\documentclass[aps,11pt]{revtex4}
\usepackage{amsmath}
\def\[{\left\lbrack}
\def\]{\right\rbrack}

\def\({\left(}
\def\){\right)}
\newcommand{\be}{\begin{equation}}
\newcommand{\ee}{\end{equation}}
\newcommand{\ea}{\end{eqnarray}}
\newcommand{\ba}{\begin{eqnarray}}

\begin{document}

\title{Supersymmetrization of the Radiation Damping}

\author{A.C.R. Mendes, C. Neves, W. Oliveira and F.I. Takakura }
\thanks{\noindent e-mail:albert@fisica.ufjf.br, cneves@fisica.ufjf.br,\\ wilson@fisica.ufjf.br, takakura@fisica.ufjf.br}
\affiliation{Departamento de F\'{\i}sica, Universidade Federal de
Juiz de Fora, 36036-330, Juiz de Fora, MG, Brasil}

\begin{abstract}
We construct a supersymmetrized version of the model to the radiation damping \cite{03} introduced by the present authors \cite{ACWF}. We dicuss its symmetries and the corresponding conserved Noether charges. It is shown this supersymmetric version  provides a supersymmetric generalization of the Galilei algebra obtained in \cite{ACWF}. We have shown that the supersymmetric action can be splited into dynamically independent external and internal sectors.
\end{abstract}

\maketitle

\section{Introduction}\label{intr}
A fundamental property of all charge particles is that electromagnetic energy is radiated whenever they are accelerated. The recoil momentum of the photons emitted during this process is equivalent to a reaction force corresponding to the self-interaction of the particle with its own electromagnetic field, which originates the Radiation Damping \cite{03}. 

The process of radiation damping is important in many areas of
electron accelerator operation \cite{04}, like in recent
experiments with intense-laser relativistic-electron scattering at
lasers frequencies and field strengths where radiation reaction
forces begin to become significant \cite{13,14}.

In \cite{ACWF} the present authors presented a new approach in the study of the radiation damping, introducing a Lagrangian formalism to the model in D=2+1 dimensions given by
\be\label{01}
L={1\over2}m g_{ij}{\dot x}_i {\dot x}_j -{\gamma \over 2}{\dot x}_i {\ddot x}_j, \;\; i,j=1,2,
\ee
where $\epsilon_{ij}$ is the Levi-Civita antisymmetric metric, $g_{ij}$ is the pseudo-euclidian metric given by
\be\label{02}
g= \left( \begin{array}{cc}
1 & 0 \\
0 & -1
\end{array}
\right),
\ee
and where, as will be the case throughout the paper, the Einstein convention on the summation of repeated indices is employed. This formalism represent a new scenario in the study of this very interesting system. The Lagrangian (\ref{01}) describes, in the hyperbolic plane, the dissipative system of a charge interacting with its own radiation, where the 2-system represents the reservoir or heat bath coupled to the 1-system   \cite{ACWF}. The model (\ref{01}) was shown to have the (2+1) Galilean symmetry and  the dynamical group structure associated to that system is the $SU(1,1)$ \cite{ACWF}. Note that, this Lagrangian is similar to the one discussed by Lukierski et al \cite{06}, but in this case we have a pseudo euclidian metric and the radiation damping constant $\gamma$ is the coupling constant of a Chern-Simons-like term.

We study in this paper a supersymmetrized version of the model (\ref{01}), where we employ a supersymmetric enlargement of the Galilei algebra obtained in \cite{ACWF} and the supersymmetries of the model are determined. We also introduce the split into ``external'' and ``internal'' degrees of freedom of the supersymmetric model (\ref{01}) in terms of new variables, where the radiation damping constant introduces noncommutativity in the coordinate sector. The dynamic splits into the decoupled sum of the dynamics in the physical sector  and in the auxiliary sector \cite{ACWF}.

The paper is organized as follows. In the section \ref{super}, we introduce the supersymmetric model and their canonical structure. In section \ref{noether},  the Noether charges associated with the symmetries are obtained. In section \ref{intext}, we shown that the supersymmetric Lagrangian can be splited into ``external'' and ``internal'' degrees of freedom and obtain the symmetries associated with each sector. In the final section we present our concluding remarks and final comments.

\section{The supersymmetric model and their canonical structure}\label{super}
The supersymmetric extension of (\ref{01}) is obtained introducing two-dimensional free-fermion term in (\ref{01}):
\be\label{03}
\bar{L}=L + i{m\over 2}g_{ij}\psi_i \dot \psi_j +i{\gamma \over 2}\epsilon_{ij}\dot \psi_i \dot \psi_j ,\;\; \psi_i , \;\; i=1,2 \;\;(\rm{Grassmannian})
\ee 
or
\be\label{04}
\bar{L}={1\over2}m g_{ij}{\dot x}_i {\dot x}_j -{\gamma \over 2}{\dot x}_i {\ddot x}_j+ i{m\over 2}g_{ij}\psi_i \dot \psi_j +i{\gamma \over 2}\epsilon_{ij}\dot \psi_i \dot \psi_j .
\ee 

The Euler-Lagrange equations are
\be\label{05}
{d\over {dt}}{{\partial \bar{L}}\over{\partial \dot x_i}}-{d^2\over {dt^2}}{{\partial \bar{L}}\over{\partial \ddot x_i }}-{{\partial \bar{L}}\over{\partial x_i}} =0 \;\;{\rm{and}}\;\; {d\over {dt}}{{\partial \bar{L}}\over{\partial \ddot \psi_i}}-{{\partial \bar{L}}\over{\partial \psi_i }}=0
\ee
given the following equations of motion for our model
\begin{subequations}\label{6}
\begin{align}
\frac{d}{dt}\( mg_{ij}\dot x_j -\gamma \epsilon_{ij}\ddot x_j \) &=0 \rightarrow mg_{ij} \ddot x_j -\gamma\epsilon_{ij}\stackrel{\ldots}x_j =0,\label{6a}\\
\frac{d}{dt} \(-i\frac{m}{2} g_{ij}\psi_j + i\gamma \epsilon_{ij} \dot \psi_j \)-i\frac{m}{2} g_{ij} \dot \psi_j &=0 \rightarrow -img_{ij}\dot \psi_j + i\gamma \epsilon_{ij} \ddot \psi_j =0 .\label{6b}
\end{align}
\end{subequations}

Note that the right derivative is employed to define the derivative in the fermionic coordinates. This convention will be used throughout the paper.

Due  to the presence of a second order time-derivatives in the Lagrangian we have to introduce three momenta:
\begin{subequations}
\begin{align}
p_i &= \frac{\partial \bar{L}}{\partial \dot x_i}-\frac{d}{dt}\frac{\partial \bar{L}}{\partial \ddot x_i}=mg_{ij}\dot x_j -\gamma \epsilon_{ij}\ddot x_j ,\label{7a}\\
\tilde p_i &=\frac{\partial \bar{L}}{\partial \ddot x_i} = \frac{\gamma}{2}\epsilon_{ij}\dot x_j ,\label{7b} \\
\pi_i &= \frac{\partial \bar{L}}{\partial \dot \psi_i}= -i\frac{m}{2}g_{ij}\psi_j +i\gamma \epsilon_{ij} \dot \psi_j .\label{7c}
\end{align}
\end{subequations}
This suggests that twelve canonical variables $\{x_i ,\dot x_i , p_i , \tilde p_i ; \psi_i ,\pi_i \}$ should be employed. However, the elements in this set of canonical variables are not independent, because our model have two constraints (see Eq.(\ref{7b})) of the second class \cite{Dirac}
\be\label{08}
\Phi_i = \dot x_i +{2\over \gamma}\epsilon_{ij}\tilde p_j .
\ee
The Hamiltonian formalism for the Lagrangian (\ref{04}) can be written in the ten-dimensional phase space $\{ x_i ;p_i ;\tilde p_i ; \psi_i ; \pi_i \}$, using the Legendre transformation,
\ba\label{09}
\bar H &=&\dot x_i p_i +\ddot x_i \tilde p_i +\dot \psi_i \pi_i -\bar L\nonumber\\
&=& H_b +H_f,
\ea
where the Hamiltonian  for the bosonic sector, $H_b$ (obtained in \cite{ACWF}) is
\be\label{10}
H_b ={{2m}\over\gamma^2}g_{ij}\tilde p_i \tilde p_j -{2\over \gamma} p_i \epsilon_{ij} \tilde p_j ,
\ee
and, for the fermionic sector is
\be\label{11}
H_f = -{i\over{2\gamma}}\epsilon_{ij}\( \pi_i + i{m\over 2}g_{il}\psi_l \) \( \pi_j +i{m\over2}g_{jk}\psi_k \).
\ee

Next, we want to investigate the canonical equations of motion and the Poisson algebra of the model. But, due to the constraints (\ref{08}) it is necessary to use the graded Poisson bracket defined as
\be\label{12}
\{A,B\}=\({{\partial A}\over{\partial x_i}}{{\partial B}\over{\partial p_i}}-{{\partial A}\over{\partial p_i}}{{\partial B}\over{\partial x_i}}\)+\( {{\partial A}\over{\partial \dot x_i}}{{\partial B}\over{\partial \tilde p_i}}-{{\partial A}\over{\partial \tilde p_i}}{{\partial B}\over{\partial \dot x_i}} \)- \( {{\partial B}\over{\partial \pi_i}}{{\partial A}\over{\partial \psi_i}}+ {{\partial B}\over{\partial \psi_i}}{{\partial A}\over{\partial \pi_i}}\),
\ee
as well as the Dirac bracket \cite{Dirac}
\be\label{13}
\{ A,B\}_D = \{ A,B \} -\{ A, \Phi_i \}C_{ij}^{-1} \{\Phi_j ,B\} ,
\ee
where $A$, $B$ can be either bosonic or fermionic valued differentiable functions of the canonical variables $\{ x_i , \dot x_i , p_i , \tilde p_i ; \psi_i , \pi_i \}$ and the matrix $C$ is defined through the relation $C_{ij} =\{ \Phi_i ,\Phi_j \}$. Hence, the non nulls canonical Poisson brackets among the canonical variables are
\be\label{14}
\{ x_i , p_j \} =\delta_{ij} ; \;\; \{ \dot x_i ,\tilde p_j \} =\delta _{ij} ; \;\; \{\psi_i ,\pi_j \} =-\delta_{ij}.
\ee
Substituting  the constraints (\ref{08}) in (\ref{13}) we have for the model,
\be\label{15}
\{ A,B\}_D = \{ A,B \} -\{ A, \Phi_i \}{\gamma \over 4}\epsilon_{ij} \{\Phi_j ,B\} .
\ee
In particular, the fundamental Poisson bracket relations are replaced by the symplectic structure depending on the choice of ten independent canonical variables. Choosing the independent variables as $y_a =\{ x_i , p_i , \tilde p_i ; \psi_i , \pi_i \}$, $a= 1 \ldots ,10$ we get
\be\label{16}
\{ y_a ,y_b \}_D =\omega_{ab} ,
\ee
where
\be\label{17}
\omega = \left( \begin{array}{ccccc}
\textbf{0} & \textbf{1}_2 & \textbf{0} & \textbf{0} & \textbf{0} \\
-\textbf{1}_2 & \textbf{0} & \textbf{0} & \textbf{0} & \textbf{0} \\
\textbf{0} & \textbf{0} & {\gamma \over 2}\epsilon & \textbf{0} & \textbf{0} \\
\textbf{0} & \textbf{0} & \textbf{0} & \textbf{0} & -\textbf{1}_2 \\
\textbf{0} & \textbf{0} & \textbf{0} & -\textbf{1}_2 & \textbf{0}
\end{array} \right),
\ee
with
\be\label{18}
\textbf{1}_2 =\left(\begin{array}{cc}
1 & 0 \\
0 & 1 
\end{array}\right),\;\; \epsilon=\left( \begin{array}{cc}
0 & 1 \\
-1 & 0 
\end{array}\right),
\ee
and \textbf{0} denotes the 2$\times$2 null matrix.

The Hamiltonian equations of motion
\be\label{19}
\dot y_a =\{ y_a , \bar{H} \}_D \ee
take the form 
\begin{subequations}
\begin{align}
\dot x_i &= \{ x_i , {\bar H} \}_D = -\frac{2}{\gamma}\epsilon_{ij} p_j ,\label{20a}\\
\dot p_i &= \{ p_i ,\bar{H}\}_D =0 , \label{20b}\\
\dot {\tilde p_i} &= \{ {\tilde p_i} , \bar{H} \}_D =\frac{m}{\gamma}g_{ij}\tilde p_j -\frac{1}{2}p_i , \label{20c}\\
\dot \psi_i &= \{ \psi_i , \bar{H} \}_D =\frac{i}{\gamma}\epsilon_{ij} \( \pi_j + i\frac{m}{2}g_{jl}\psi_l \) ,\label{20d}\\
\dot \pi_i &= \{ \pi_i  , \bar{H} \}_D =-\frac{m}{2\gamma}g_{il}\epsilon_{lj} \( \pi_j +i\frac{m}{2}g_{jk}\psi_k \),\label{20e}
\end{align}
\end{subequations}
where $\bar H$ is given by (\ref{09}).
These equations of motion are  consistent with the Euler-Lagrange equations (\ref{6}) derived from the Lagrangian. To obtain the quantized form of the canonical commutation relation (\ref{19}) as well as the Heisenberg equations of motion we perform the replacement
\be\label{21}
\{ y , y^{'} \}_D \rightarrow {1\over {i\hbar }}[ \hat{y} , \hat{y}^{'} ] , \ee
where $\hat{y}$, $\hat{y}^{'}$ denote the quantized variables.

\section{The Noether charges and their symmetries}\label{noether}

Let us consider a Lagrangian $\bar{L}( x_i ,\dot x_i , \ddot x_i ; \psi_i , \dot \psi_i )$ which depends on the first and second time derivatives. The variation of the action $S=\int dt\; \bar{L}$ under the change $x_i \rightarrow x_i + \delta x_i$ and $\psi_i \rightarrow \psi_i +\delta \psi_i$, bosonic and fermionic variables respectively, takes the form
\ba\label{22}
\delta S =\int \; \delta \bar{L} &=& \int dt \;\( \delta x_i {{\partial \bar{L}}\over{\partial x_i}} + \delta \dot x_i {{\partial \bar{L}}\over{\partial \dot x_i}} + \delta \ddot x_i {{\partial \bar{L}}\over{\partial \ddot x_i}} +\delta \psi_i {{\partial \bar{L}}\over{\partial \psi_i}} +\delta \dot \psi_i {{\partial \bar{L}}\over{\partial \dot \psi_i}} \) \nonumber\\
&=& \int dt\; {d\over{dt}}\( \delta x_i p_i + \delta \dot x_i \tilde p_i +\delta \psi_i \pi_i \),
\ea
from where we obtain the following formula for the generator
\be\label{23}
C(t)=\delta x_i p_i + \delta \dot x_i \tilde p_i +\delta \psi_i \pi_i ,
\ee
which is conserved $\( {d\over{dt}}C(t) =0 \)$.

Let us list the generators of the symmetry for the Lagrangian (\ref{04});\\

\noindent (i) space-translations: $\delta x_i =\delta_i ,\; \delta \dot x_i =0 ;\; \delta \psi_i = \bar{\delta}_i $ ($\bar{\delta}_i$ are Grassmannian), $\delta \dot \psi_i =0$, where $\delta_i$ and $\bar{\delta}_i$ are respectively the translation shifts of the bosonic and fermionic variables. So
\be\label{24}
C_t = \delta_i p_i + \bar{\delta}_i \pi_i  .
\ee
But, the Lagrangian (\ref{04}) is quasi-invariant under space-translations transformations; in fact
\be\label{25}
\delta_t \bar{L}= \bar{\delta}_i {d\over{dt}}\(i {m\over 2}g_{ij}\psi_j \).
\ee
So, the generators (\ref{24}) are not conserved. However, as the nonconservation law takes the form
\be\label{26}
{d\over{dt}}G(t) ={d\over{dt}}\Lambda(t),
\ee
we can introduce modified generators $\tilde G=G-\Lambda$, which are conserved. In this case, we derive the following conserved generator
\be\label{27}
\tilde G_t = P_i \delta_i  +\Pi_i \bar{\delta}_i 
\ee
where
\be\label{28}
P_i =p_i \;\; {\rm and }\;\; \Pi_i =\pi_i -i{m\over 2}g_{ij} \psi_j .
\ee

\noindent (ii) rotations: $\delta_r x_i =-\epsilon_{ij}x_j \phi_b ,\; \delta_r \dot x_i =-\epsilon_{ij}\dot x_j \phi_b; \; \delta_r \psi_i =-\epsilon_{ij}\psi_j \phi_f $, where $\phi_b$ and $\phi_f$ are respectively the rotation angles of the bosonic and fermionic variables, then
\be\label{29}
G_r = -\epsilon_{ij}p_i x_j \phi_b -\epsilon_{ij}p_i \dot x_j \phi_b -\epsilon_{ij}\pi_i \psi_j \phi_f .
\ee
Using the constraint equations (\ref{08}) we find that
\be\label{30}
G_r = J_b \phi_b -J_f \phi_f
\ee
where
\be\label{31}
J_b =x_i \epsilon_{ij} p_j -{2\over \gamma} \tilde p^2_i ,\;\; J_f = -\psi_i \epsilon_{ij} \pi_j  .
\ee

\noindent (iii) Galilei boosts: $\delta_{v_i} x_i = v_i t ,\; \delta_{v_i} \dot {x_i} =  v_i ; \; \delta_{v_i} \psi =0$ (Note that the Grassmannian variables do not transform under Galilei boosts), then
\be\label{32}
G_{v_i} =   v_i p_i  t +v_i \tilde p_i   .
\ee
But, the Lagrangian (\ref{04}) remains invariant up to a total time-derivative under Galilei boosts transformation:
\be\label{33}
\delta_{v_i} \bar{L} = {d\over{dt}} \( mg_{ij} -{\gamma \over 2}\epsilon_{ij} \dot x_i \) v_i
.
\ee
So, using (\ref{26}) we derive the following conserved generator
\be\label{34}
\tilde G_{v_i} = \tilde B_i v_i
\ee
where
\be\label{35}
\tilde B_i =p_i  t + \tilde p_i  -mg_{ij} -{\gamma \over 2}\epsilon_{ij} \dot x_i.
\ee
Using the constraint equations (\ref{08}), we get the following conserved generator
\be\label{36}
\tilde B_i = p_i t -mg_{ij} x_j + 2\tilde p_i.
\ee

\noindent (iv) time: $\delta_{\tau}t=\tau$ ($\tau$ is the translation shift of the time variables)

As the Lagrangian of the model (\ref{04}) does not explicitly depend on  time, the conserved quantity corresponding to the time-translation is given by the Hamiltonian (\ref{09}).

\noindent (v) supersymmetry: $\delta_Q x_i = i\epsilon \psi_i ,\; \delta_Q \dot x_i =i\epsilon \dot \psi_i ; \; \delta_Q \psi _i =-\epsilon \dot x_i  $ ($\epsilon$ is an infinitesimal Grassmannian parameter), then
\be\label{37}
G_Q = i\epsilon \psi_i p_i +i\epsilon \dot \psi_i \tilde p_i -\epsilon \dot x_i \pi_i .
\ee
However, the Lagrangian (\ref{04}) remains also invariant up to a total time-derivative under supersymmetric transformation, so 
\be\label{38}
\delta_Q \bar{L} =\epsilon{d\over{dt}}\( i{m\over 2}g_{ij} \dot x_i \psi_j -i {\gamma \over 2}\epsilon_{ij} \dot x_i \dot \psi_j \).
\ee
Using the relation (\ref{26}) and introducing the constraint equations (\ref{08}), we get the conserved generator
\be\label{39}
\tilde G_Q = Q \epsilon ,
\ee
then
\be\label{40}
Q =i p_i \psi_i -{2\over\gamma}\epsilon_{ij} \tilde p_i \( \pi_j + i{m\over 2}g_{jk}\psi_k \),
\ee
where the Poisson algebra of this supercharge is given by
\be\label{40.1}
-{i\over2}\{ Q,Q \}_D  =H_b + H_f =\bar{H}
\ee
which is the Hamiltonian of the model (see (\ref{09})).

\section{Supersymmetry in External and Internal sectors}\label{intext}

Now, we will show that the supersymmetric Lagrangian (\ref{04}) can be splited into ``external'' and ``internal'' degrees of freedom dinamically independent as in \cite{ACWF}. To this end, following Faddeev-Jackiw's method of describing  Lagrangians with higher-order derivatives \cite{08} we describe, equivalently, the action (\ref{04}) as
\be\label{41}
\bar{L}^{(0)} = {1\over 2}g_{ij}v_i v_j -{\gamma\over 2}\epsilon_{ij}v_i \dot v_j + {i\over 2}g_{ij} \psi_i \dot \psi_j +i\gamma\epsilon_{ij} \dot \psi_i \rho_j -i{\gamma\over 2}\epsilon_{ij}\rho_i \rho_j + p_i (\dot x_i -v_i ),
\ee
where the field equation for $\rho_i$ is purely algebraic and where $\rho_i$ is fermionic (we put for simplicity  $m=1$).

Now, analysing the first-order Lagrangian $\bar{L}^{(0)}$, equation (\ref{41}), from the symplectic point of view \cite{08}, the Dirac brackets among the phase space variables $\xi_a =\{x_i ,p_i , v_i ; \psi_i , \rho_i \}$ are
\be\label{42}
\{ \xi_a ,\xi_b \}_D=f^{-1}_{ab},
\ee
where $f^{-1}_{ab}$ is the inverse of the symplectic matrix
\be\label{43}
f=\left( \begin{array}{ccccc}
\textbf{0} & -\textbf{1}_2 & \textbf{0} & \textbf{0} & \textbf{0} \\
\textbf{1}_2 & \textbf{0} & \textbf{0} & \textbf{0} & \textbf{0} \\
\textbf{0} & \textbf{0} & \gamma\epsilon & \textbf{0} & \textbf{0} \\
\textbf{0} & \textbf{0} & \textbf{0} & -ig & i\gamma \epsilon \\
\textbf{0} & \textbf{0} & \textbf{0} & i\gamma \epsilon & \textbf{0} 
\end{array}\right).
\ee
So, for the fermionic sector of the Lagrangian (\ref{41}) one obtains 
\be\label{44}
\{ \psi_i ,\psi_j \}_D =0; \;\; \{ \psi_i ,\rho_j \}_D=-{i\over\gamma}\epsilon_{ij} ;\;\; \{\rho_i ,\rho_j \}_D ={i\over\gamma}g_{ij}.
\ee
The Dirac brackets for the bosonic sector are the same as the ones obtained in Section \ref{super}.

In order to split (\ref{41}) into external and internal sector, the variables introduced in \cite{Horvathy,Lukierski} are modified as
\ba\label{45}
{\cal Q}_i &=&\gamma \(g_{ij} v_j -p_i \), \nonumber\\
X_i &=& x_i + \epsilon_{ij}{\cal Q}_j , \\
P_i &=& p_i , \nonumber\\
\tilde \psi_i &=& \psi_i - \gamma g_{ik} \epsilon_{kj}\rho_j , \nonumber
\ea
giving the set of canonical Poisson brackets:
\be\label{46}
\{ X_i ,X_j \}_D = -\gamma\epsilon_{ij}, \;\; \{ P_i ,P_j \}_D =0,\;\; \{ X_i ,P_j \}_D =\delta_{ij}, \;\; \{ {\cal Q}_i ,{\cal Q}_j \}_D =-\gamma\epsilon_{ij},
\ee
for the bosonic sector.For the  fermionic sector, the new fermionic variables satisfy the following Poisson brackets algebra
\be\label{46.a}
\{ \tilde \psi_i ,\tilde \psi_j \}_D =ig_{ij}, \;\; \{ \tilde \psi_i ,\rho_j \}_D =0.
\ee

Now, substituting (\ref{45}) in (\ref{41}) the action takes the form
\be\label{47}
\bar{L}^{(0)} = \bar{L}^{(0)}_{\rm{ext}}+  \bar{L}^{(0)}_{\rm{int}}
\ee
where
\be\label{48}
\bar{L}^{(0)}_{\rm{ext}} = P_i \dot X_i + {\gamma\over2}\epsilon_{ij}P_i \dot P_j -{1\over 2}g_{ij}P_i P_j + {i\over 2} g_{ij} \tilde \psi_i \dot{\tilde\psi}_j ,
\ee
\be\label{49}
\bar{L}^{(0)}_{\rm{int}} = {1\over{2\gamma}}\epsilon_{ij}{\cal Q}_i \dot{\cal Q}_j +{1\over{2\gamma^2}}g_{ij}{\cal Q}_i {\cal Q}_j + i{\gamma^2 \over 2}g_{ij} \rho_i \dot \rho_j -i{\gamma\over 2}\epsilon_{ij}\rho_i \rho_j .
\ee
We see that our Lagrangian separates into two disconnected parts describing the ``external'' and ``internal'' degrees of freedom both describing supersymmetric models. Note that, while original coordinates commute, $\{ x_i ,x_j \}=0$, both the ``external'' and ``internal'' positions, $X_i$ and ${\cal Q}_i$, respectively, are non-commuting (see (\ref{46})).

Theses actions are invariant under the following set of supersymmetry transformations:

\noindent (i) for the external sector, equation (\ref{48}), we get
\ba\label{50}
\delta_Q P_i &=&0,\nonumber\\
\delta_Q \psi_i &=&-\epsilon P_i ,\\
\delta_Q X_i &=& ig_{ij}\epsilon \tilde \psi_j. \nonumber 
\ea 
\noindent (ii) for the internal sector, equation (\ref{49}), we get
\ba\label{50.1}
\delta_Q {\cal Q}_i &=& i\gamma \epsilon \epsilon_{ij} \rho_j ,\nonumber\\
\delta_Q \rho_i &=& {1\over \gamma^2}\epsilon g_{ij} {\cal Q}_j,
\ea
where $\epsilon$ is a constant Grassmann number.

The supercharge corresponding to (\ref{48}), generator for the transformations (\ref{50}), is given by the formula
\be\label{51}
Q_{\rm{ext}} =J^0_{\rm{ext}}={{\delta\bar{L}^{(0)}_{\rm{ext}}}\over{\delta \dot \varsigma_i}}{{\delta \varsigma_i}\over{\delta\phi}}=ig_{ij}\tilde \psi_i P_j,
\ee
where $J^0_{\rm{ext}}$ is the conserved current and $\delta \phi$ is an infinitesimal parameter. The external Hamiltonian can be obtained consistently as
\be\label{52}
\bar{H}^{(0)}_{\rm{ext}}=-{i\over2}\{ Q_{\rm{ext}}, Q_{\rm{ext}} \}_D ={1\over2}g_{ij}P_i P_j .
\ee 
Similarly, the supercharge corresponding to (\ref{49}), generator of the transformations (\ref{50.1}), is given by
\be\label{53}
Q_{\rm{int}} =J^0_{\rm{int}}={{\delta\bar{L}^{(0)}_{\rm{int}}}\over{\delta \dot \varsigma_i}}{{\delta \varsigma_i}\over{\delta\phi}}=-{i\over 2}{\cal Q}_i \rho_i,
\ee
and our internal Hamiltonian is given by
\be\label{54}
\bar{H}^{(0)}_{\rm{int}} =-{i\over 2}\{ Q_{\rm{int}} ,Q_{\rm{int}} \}_D =-{i\over {2\gamma^2}}g_{ij}{\cal Q}_i {\cal Q}_j +i{\gamma \over 2} \epsilon_{ij} \rho_i \rho_j .
\ee

\section{Concluding remarks}

In this paper we have presented a complete formulation of a supersymmetrized version of the model (\ref{01}).  We have started with the construction of the supersymmetric model (\ref{04}) then, using the Dirac formalism for constrained Hamiltonian systems, the equations of motion and the canonical structure of the supersymmetric model are presented.

Next, we construct the (2+1) dimensional Galilean supersymmetry and supersymmetry transformations of the variables appearing in the Lagrangian. Using Noether's procedure, we construct the conserved quantities associated to these symmetries, such as supercharges (\ref{40}), which are the generators of supersymmetry transformations.

Finally, introducing non-commutative coordinates (\ref{45}) and using the Faddeev-Jackiw's method, we see therefore that the supersymmetric action (\ref{04}) can be split into dynamically independent external and internal sectors (see (\ref{48})-(\ref{49})). As we have shown, the external (\ref{48}) and internal (\ref{49}) sector remains invariant under the supersymmetryc transformation (\ref{50}) and (\ref{50.1}) respectively, and the associated supercharge is constructed.

\section{ Acknowledgments}
This work is supported in part by FAPEMIG and CNPq, Brazilian
Research Agencies. In particular, ACRM and WO would like to acknowledge
the CNPq and CN, WO and FIT would like to acknowledge the FAPEMIG.


\begin{thebibliography} {99}

\bibitem{03}W. Heitler, {\it The Quantum Theory of Radiation} (Dover, 1970), 3nd ed. ; J.D. Jackson, {\it Classical
Electrodynamics} (Wiley, New York, 1975), 2nd ed., Chaps. 14,17; A.C.R. Mendes and F.I. Takakura, Phys. Rev E {\bf 64} (2001), 056501.
\bibitem{ACWF}A.C.R. Mendes, C. Neves, W. Oliveira and F.I. Takakura: ``A new approach of canonical quantization of the radiation damping'', UFJF (2004); hep-th/0503135.
\bibitem{04}R.P. Walker,  ``Radiation Damping'', {\it Proceeding,
General Accelerator Physics}, CERN Genebra - CERN 91-04, 116-135
(1990).
\bibitem{13}F.V. Hartemann and A.K. Kerman, Phys. Rev. Lett. {\bf
76} (1996), 624.
\bibitem{14}C. Bula, K.T. McDonald, E.J. Prebys, C. Bamber, S. Boege,
T. Kotseroglou, A.C. Melissinos, D.D. Meyerhofer, W. Ragg, D.L.
Burke, R.C. Field, G. Horton-Smith, A.C. Odian, J.E. Spencer, D.
Walz, S.C. Berridge, W.M. Bugg, K. Shmakov and A.W. Weidemann,
Phys. Rev. Lett. {\bf 76} (1996), 3116.
\bibitem{06} J. Lukierski, P.C. Stichel and W.J. Zakrzewski, Ann.
Phys. {\bf 260} (1997), 224.
\bibitem{Dirac}P.A.M. Dirac, {\it Lectures on Quantum Mechanics}, Dover Publications, 2001.
\bibitem{08}L. Faddeev and R. Jackiw, Phys. Rev. Lett. {\bf 60}
(1988), 1692;\\
J. Barcelos Neto and C. Wotzasek, Mod. Phys. Lett. {\bf A7} (1992) 1737; Int. J.
Mod. Phys. {\bf A7} (1992) 4981 .
\bibitem{Horvathy}P.A. Horvathy and M.S. Plyushchay, JHEP {\bf 06}, 033 (2002).
\bibitem{Lukierski}J. Lukierski, P.C. Stichel and W.J. Zakrzewski, hep-th/0208200.


\end{thebibliography}
\end{document}